%% file: main.tex
\documentclass[runningheads]{llncs}
\usepackage{graphicx}
\usepackage{epstopdf}
\usepackage{multirow}
\graphicspath{{text/figs/}}
\begin{document}
\title{Pandemic Pulse: Unraveling and Modeling Social Signals during the COVID-19 Pandemic}

\titlerunning{Pandemic Pulse}

\author{Steven J. Krieg\thanks{Both authors contributed equally to this work.} \and
Jennifer J. Schnur{*} \and 
Jermaine D. Marshall \and
Matthew M. Schoenbauer \and
Nitesh V. Chawla }
\authorrunning{Krieg, Schnur, et. al.}
\institute{Center for Network and Data Science, University of Notre Dame, Notre Dame, IN 46556, USA \\
\email{\{skrieg,jschnur,mschoenb,jmarsha5,nchawla\}@nd.edu} \\
\url{https://cnds.nd.edu/} }
\maketitle              
\begin{abstract}
We present and begin to explore a collection of social data that represents part of the COVID-19 pandemic's effects on the United States. This data is collected from a range of sources and includes longitudinal trends of news topics, social distancing behaviors, community mobility changes, web searches, and more. This multimodal effort enables new opportunities for analyzing the impacts such a pandemic has on the pulse of society. Our preliminary results show that the number of COVID-19-related news articles published immediately after the World Health Organization declared the pandemic on March 11, and that since that time have steadily decreased---regardless of changes in the number of cases or public policies. Additionally, we found that politically moderate and scientifically-grounded sources have, relative to baselines measured before the beginning of the pandemic, published a lower proportion of COVID-19 news articles than more politically extreme sources. We suggest that further analysis of these multimodal signals could produce meaningful social insights and present an interactive dashboard to aid further exploration.\footnote{http://cnds.nd.edu/pandemicpulse}

\keywords{covid-19 \and coronavirus\and social signals \and news data \and social distancing \and mobility trends \and demographics \and web searches}
\end{abstract}

\input{text/intro}
\input{text/methods}
\input{text/results}
\input{text/conclusion}

\bibliographystyle{splncs04}
\bibliography{sources.bib}
\end{document}

%% file: text/intro.tex
\section{Introduction} \label{sec:intro}
The COVID-19 pandemic has disrupted the rhythms of society in unprecedented ways and at an unparalleled scale. In this work, we present and begin to explore a collection of social signals that represent part of the social pulse of the United States. These signals include COVID-19 case data, demographic data, longitudinal news and web search trends, media bias data, and mobility reports. As a doctor studies a patient's vitals to aid in identifying a diagnosis and prescribing treatment, we aim to unravel and model these signals to inform our understanding of broad effects of the COVID-19 pandemic on the spread of information, social behaviors, and more. To aid in further exploration, we published an interactive dashboard alongside this paper.$^1$

The rest of the paper proceeds as follows: in Section \ref{sec:methods} we describe data collection and preprocessing, in Section \ref{sec:results} we present the results of preliminary analysis of news signals, and in Section \ref{sec:conclusion} we discuss opportunities for future work.

%% file: text/methods.tex
\section{Data} \label{sec:methods}
We collected COVID-19 case data from Johns Hopkins Univerisity \cite{johnshopkins}, news data from the Global Database of Events, Language, and Tone (GDELT) \cite{leetaru2013gdelt}, web search data from Google trends, media bias labels from Media Bias/Fact Check \cite{mediabiasfactcheck} and AllSides \cite{allsides}, social distancing data from Unacast \cite{unacast}, and demographic data from the Center for Disease Control and Prevention \cite{hypertension2015,diabetesdata_2015,cancerdata,kff_2019,laborstatistics_2020,censusbureau_2020}. In the following sections, we detail our methods for collection and analysis.

\subsection{COVID-19 Case Data} \label{sec:covid19cases}
Johns Hopkins University (JHU) has created a repository for COVID-19 case data that combines information from the World Health Organization (WHO) and a number of other global and national sources \cite{johnshopkins}. We use this data from JHU to report the number of new cases and new deaths by location and date.

\subsection{United States Demographics}
In order to represent demographic information as well as risk factors based on individual states, we collected data from various sources including the Center for Disease Control, United States Census Bureau, and the Bureau of Labor Statistics. This data enables us to explore correlations between demographic information for locations and other data, such as searching for relationships between locations with higher rates of COVID-19 deaths. The demographic data we collected includes heart disease hospitalization rate, cancer rate, population age, hypertension and stroke rates, obesity, walk scores, eating habits (i.e. veggie intake), ethnicity and smoking habits. After collecting all variables for each state, we performed normal preprocessing and cleaning steps: noise removal, aggregation, and conversion to percentages.

\subsection{News Data} \label{sec:news}
\subsubsection{COVID-19 Articles} \label{sec:covid19articles}
The Global Database of Events, Language, and Tone (GDELT) monitors worldwide print, broadcast, and online news in over 100 languages \cite{leetaru2013gdelt}. For each article published, GDELT adds to its Global Knowledge Graph (GKG) a record that contains a variety of metadata including geographical references, textual themes, and sentiment scores.\footnote{While the GKG monitors other news formats, the vast majority of its COVID-19-related records represent textual pieces. We therefore use the term ``article'' to refer generally to any record stored in the GKG.} The GKG processes several terabytes of data every year, making it a rich source of longitudinal news data. We created a corpus of COVID-19 news by extracting from the GKG any record that met at least one of the criteria listed in Table \ref{tab:covidcorpus}. We also removed duplicate articles, which we defined as those with a non-unique combination of publisher and title.

\begin{table}[h]
    \centering
    \begin{tabular}{|l|l|l|}
        \hline
        \textbf{Description} & \textbf{GKG Column(s)} & \textbf{Possible Values} \\ \hline 
        \multirow{6}{*}{Article title or URL contains} & \multirow{2}{*}{} & coronavirus \\
        & & covid \\
        & Title & 2019-ncov \\
        & DocumentIdentifier & ncov-2019 \\
        & \multirow{2}{*}{} & ncov2019 \\
        & & sars-cov-2 \\ \hline
        \multirow{5}{*}{\shortstack[l]{Article text includes a \\ 
                                        reference  to the COVID-19 \\ 
                                        virus, COVID-19 cases,  \\ 
                                        pandemic, or a related term}} & \multirow{5}{*}{Themes} & WB\_2167\_PANDEMICS \\
        & & HEALTH\_PANDEMIC \\
        & & *\_CORONAVIRUS \\
        & & *\_CORONAVIRUSES \\
        & & *\_CORONAVIRUS\_INFECTIONS \\ \hline
        \end{tabular}
    \caption{Criteria used to determine whether an article from the GKG should be included in the COVID-19 corpus. The * character represents the prefix ``TAX\_DISEASE''.}
    \label{tab:covidcorpus}
\end{table}

\subsubsection{Media Bias Data} \label{sec:mediabias}
We used two independent sources for labeling the political bias of news sources: Media Bias/Fact Check (MBFC) and AllSides. MBFC is an independent online media outlet that evaluates news sources on their political bias and the factuality of their publications \cite{mediabiasfactcheck}. AllSides \cite{allsides} takes a similar task, but incorporates surveys, reviews, and additional data into their evaluation process. Both have been utilized in recent works on media bias detection \cite{ribeiro2018media,yu2019experiments,hamborg2019automated}. Table \ref{tab:mediabias} lists the possible ratings given by each organizaztion.

\begin{table}[ht]
    \centering
    \begin{tabular}{|l|l|} \hline
        \multicolumn{2}{|c|}{\textbf{Possible Bias Ratings}} \\
        \textbf{Media Bias/Fact Check} & \textbf{AllSides} \\ \hline
        Left & Left \\
        Left-center & Left-center \\
        Least Biased & Least Biased \\
        Right-center & Right-center \\
        Right & Right \\ 
        Scientific & Mixed \\ 
        Questionable Sources & \\
        Conspiracy-pseudoscience & \\ \hline
    \end{tabular}
    \caption{List of possible ratings assigned to news sources by Media Bias/Fact Check and AllSides.}
    \label{tab:mediabias}
\end{table}

We utilize MBFC as our primary source and AllSides as supplementary. We prefer MBFC for the following reasons:
\begin{enumerate}
    \item MBFC's evaluation methodology is explained in more detail, and thus more transparent.
    \item MBFC includes a ``Scientific'' category, which we found to be a helpful addition. Most of MBFC's Scientific sources were labeled ``Least Biased'' by AllSides.
    \item MBFC includes a ``Questionable Sources'' category. While this is comprised largely of extreme right sources, it also contains many extreme left sources. We found it helpful to separate these extremes from regular right and left-leaning sources.
\end{enumerate}

\subsection{Social Distancing \& Mobility Data}

\subsubsection{Unacast Social Distancing Data}
Unacast provides social distancing scores for U.S. states and counties based on cell phone GPS data \cite{unacast}. From this data set, retrieved the Daily Distance Reduction score for all states since February 24. This feature measures the change between the average distance traveled per device for each day and the average distance traveled on the same weekday during the four weeks prior to the COVID-19 outbreak in the U.S. (February 10-March 8). Based on this percent change, each state is given a letter grade on each day according to the following rules \cite{unacast_methodology}:
\begin{itemize}
    \item \textbf{A}: $  > 70\%$ decrease
    \item \textbf{B}: $ 55\mathrm{-}70\%$ decrease
    \item \textbf{C}: $ 40\mathrm{-}55\%$ decrease
    \item \textbf{D}: $ 25\mathrm{-}40\%$ decrease
    \item \textbf{F}: $ < 25\%$ decrease or increase. 
\end{itemize}

\subsubsection{Google Community Mobility Reports}

The publicly available global Google Mobility Report \cite{GoogleMobility} describes longitudinal changes in population movement trends over the course of the COVID-19 outbreak. These movement trends are divided into categories for retail and recreation, groceries and pharmacies, parks, transit stations, workplaces, and residential. For each category, the report provides the percent change in visitation or time spent in places of that category relative to a baseline, which is computed as the median value for each weekday from the 5‑week period January 3, 2020-February 6, 2020. The data is aggregated from anonymized users who have opted in to sharing their location history in Google Maps. 

\subsection{Google Search Trends}
Using our collection of COVID-19-related news, we first extracted a set of keywords by tokenizing and lemmatizing the titles of each news article. Next, we retrieved the $1000$ most frequently mentioned terms, the first 10 of which are reported in Table \ref{tab:keywords}.

\begin{table}[ht]
    \centering
    \begin{tabular}{|lr|} \hline
        \textbf{Keyword} & \textbf{News Mentions} \\ \hline
        coronavirus & 2,267,125 \\
        covid-19 & 1,291,640 \\
        news & 566,882 \\
        new & 492,032 \\
        case & 483,070 \\
        virus & 457,932 \\
        pandemic & 355,369 \\
        say & 348,860 \\
        death & 314,209 \\
        trump & 263,341 \\ \hline
    \end{tabular}
    \caption{The 10 keywords that appear most frequently in the titles of COVID-19-related news articles.}
    \label{tab:keywords}
\end{table}

 We then scraped Google Trends \cite{GoogleTrends} for the longitudinal ``Interest over Time'' of each keyword from January 1 to May 31, 2020, in each U.S. state. For each keyword, Trends measures web search popularity by taking an anonymized sample of Google searches and dividing the total count of searches containing the given keyword by the total searches associated with a particular location and time range. This value is normalized between 0 and 100 in order to represent search interest relative to the given state and time, where 100 represents peak popularity for the term and 0 represents a lack of available data for the given term.

%% file: text/results.tex
\section{Prelimary Analysis} \label{sec:results}
\subsection{Quantity of News} \label{sec:newsresults}
\begin{figure}
\includegraphics[width=\textwidth]{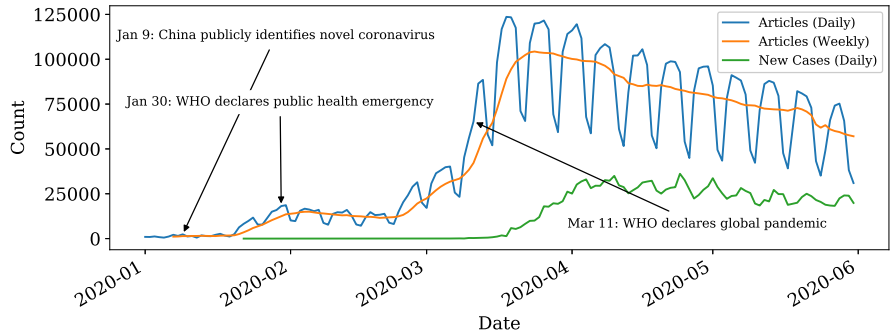}
\caption{The number of COVID-19 related articles extracted from the GKG, measured daily and weekly from Jan. 1 through May 31, 2020, plotted alongside the number of new COVID-19 cases reported in the U.S.} \label{fig:articlecounts}
\end{figure}

Through May 31, 2020, we have extracted data on over 7.6 million news articles related to the COVID-19 pandemic. Figure \ref{fig:articlecounts} shows the daily and weekly article counts from Jan. 1 through May 31, 2020. The daily oscillation represents a consistent pattern that fewer articles are published on Saturdays and Sundays. The weekly coverage increased at the end of January, around when the first case was confirmed in the United States (Jan. 20) and the Chinese authorities quarantined the city of Wuhan (Jan. 23). A local peak of 18,636 articles were published on Jan. 31, the day after the World Health Organization (WHO) declared a public health emergency. However, average weekly coverage slowly declined until the last week of February, when cases surged in Italy and Iran. At this point news coverage surged through the first reported death in the United States (Feb. 29) and the WHO's declaration of a global pandemic (Mar. 11) to a global peak of 123,623 articles (Mar. 18). Since then, coverage has decreased steadily, even as new cases reached their global peak of 36,163 (Apr. 24). Even after the number of new cases has begun to decrease, the news coverage has continued to decrease at a faster rate. This suggests that, on a broad scale, news sources were most interested in reporting the novel events surrounding the beginning of the pandemic.

\subsection{News Coverage by Political Bias} \label{sec:biasresults}
\begin{figure}
\includegraphics[width=\textwidth]{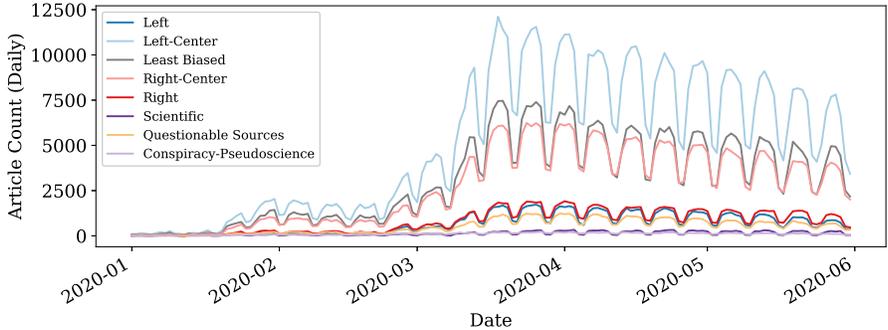}
\caption{The number of COVID-19 related articles extracted from the GKG and grouped by source bias, measured daily from Jan. 1 through May 31, 2020.} \label{fig:biascounts}
\end{figure}

Of the 7.6 million articles extracted from the GKG, just under 2 million were published by the sources evaluated for bias by MBFC or AllSides. Figure \ref{fig:biascounts} shows the daily count of articles published by each bias category, each of which follow a similar trend to the total article count. This is corroborated by Pearson tests performed with respect to the normalized distribution of articles from all sources (Figure \ref{fig:articlecounts}), which report correlation coefficients $\geq0.99$ for each bias category except ``Scientific'' and ``Conspiracy-pseudoscience,'' which report coefficients of $0.91$ and $0.92$, respectively.

\begin{figure}
\includegraphics[width=\textwidth]{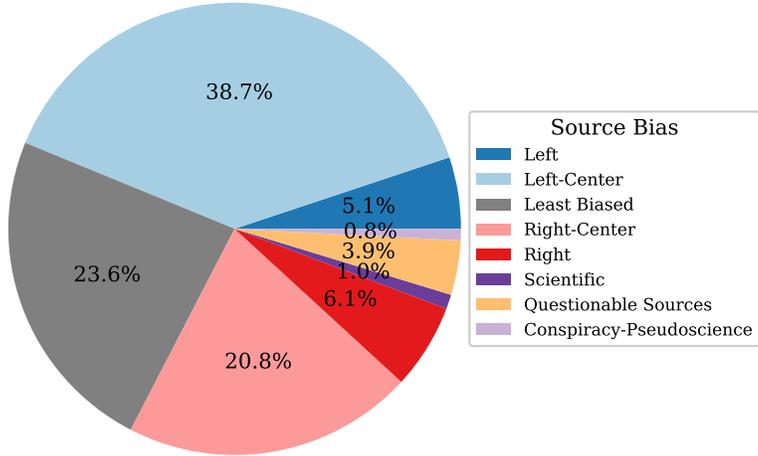}
\caption{The representation of each bias category in COVID-19-related news, measured as a percentage of all articles.} \label{fig:biaspercentage}
\end{figure}

The lower correlation of the distribution of articles published by these two bias categories may be attributable to noise. As Figure \ref{fig:biaspercentage} shows, both Scientific and Conspiracy-pseudoscience represent only a small percentage of the collection of COVID-19-related articles. However, we found that the representation of articles published by Scientific sources, when measured as a percentage of total published news, is significantly lower (0.68x) for COVID-19-related news when compared to a baseline of all 2019 articles, of which ``Scientific'' sources accounted for 1.5\% of the records. However, some bias categories have increased representation in COVID-19-related news: Right sources increased their representation by 1.15x, Right-center sources by 1.07x, and Left sources by 1.04x. This could be due to the fact that sources with stronger political biases are publishing more news than their baseline, that more moderate and scientific sources are publishing less, or a combination of the two.

\begin{figure}
\includegraphics[width=\textwidth]{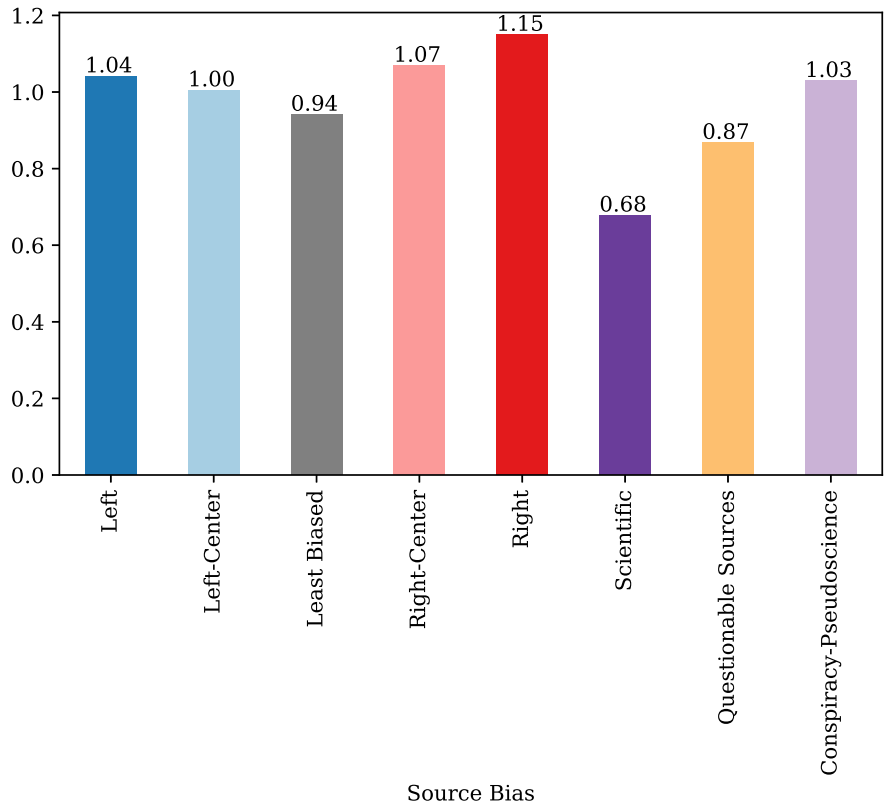}
\caption{The increase of each bias category's representation in COVID-19-related news, measured as a ratio of the percentage representation of articles in the COVID-19-related news against a baseline of all 2019 news.} \label{fig:biasdeltas}
\end{figure}

%% file: text/conclusion.tex
\section{Future Work} \label{sec:conclusion}
By aggregating multimodal data from many sources that represent a variety of social signals in the United States, we have begun to explore the effects of the COVID-19 pandemic on the pulse of U.S. society. Our current data includes COVID-19 case data, demographic data, longitudinal news and web search trends, media bias data, and mobility reports, but there are many other types of social signals that could be studied in order to better understand and model the effects of the pandemic. These could include social media trends, economic patterns, and additional healthcare data. In beginning to explore this data, we analyzed the quantity of news coverage, and showed that the amount of COVID-19-related news peaked just after the announcement of the pandemic, after which it steadily decreased. We additionally explored media bias and demonstrated that, with respect to quantity, all groups of political biases published news in a similar pattern, and that more scientific sources have significantly less representation in the COVID-19-related news when compared to their pre-pandemic baseline. There are many opportunities to examine other relationships between signals, such as the influence of news on social distancing and web searches, correlations between web searches and news topics, and differences of these effects between locations and demographics. We additionally hope to extend this data and work beyond the United States to understand the effects of the COVID-19 pandemic around the world.